# Pressure-Induced Large Volume Collapse, Plane-to-Chain, Insulator to Metal Transition in CaMn$_2$Bi$_2$

*Xin Gui,[a] Gregory J. Finkelstein,[b] Keyu Chen,[cd] Tommy Yong,[b] Przemyslaw Dera,[b] Jinguang Cheng,[cde]\* Weiwei Xie [a]\**

[a] Department of Chemistry, Louisiana State University, Baton Rouge, LA, USA 70803
[b] Hawai'i Institute of Geophysics and Planetology, University of Hawai'i at Manoa, Honolulu, HI, USA 96822
[c] Beijing National Laboratory for Condensed Matter Physics and Institute of Physics, Chinese Academy of Sciences, Beijing, China 100190
[d] School of Physical Sciences, University of Chinese Academy of Sciences, Beijing, China 100190
[e] Songshan Lake Materials Laboratory, Dongguan, Guangdong, China 523808

## ABSTRACT

In-situ high pressure single crystal X-ray diffraction study reveals that the quantum material CaMn$_2$Bi$_2$ undergoes a unique plane to chain structural transition between 2 and 3 GPa, accompanied by a large volume collapse. CaMn$_2$Bi$_2$ displays a new structure type above 2.3 GPa, with the puckered Mn honeycomb lattice of the trigonal ambient-pressure structure converting to one-dimensional (1D) zigzag chains in the high-pressure monoclinic structure. Single crystal measurements reveal that the pressure-induced structural transformation is accompanied by a dramatic two order of magnitude drop of resistivity; although the ambient pressure phase displays semiconducting behavior at low temperatures, metallic temperature dependent resistivity is observed for the high pressure phase, as, surprisingly, are two resistivity anomalies with opposite pressure dependences. Based on the electronic structure calculations, we hypothesized that the newly emerged electronic state under high pressure is associated with a Fermi surface instability of the quasi-1D Mn chains, while we infer that the other is a magnetic transition. Assessment of the total energies for hypothetical magnetic structures for high pressure CaMn$_2$Bi$_2$ indicates that ferrimagnetism is thermodynamically favored.

The design and discovery of new quantum materials and advances in our ability to predict, detect, and exploit their properties will accelerate the development of future technologies.[1-3] In solid-state quantum materials, conventional electronic and spin ordering can sometimes be broken down by quantum fluctuations, and consequently entirely new forms of electronic or spin ordering can be observed.[4-6] Therefore, there is an urgent need to understand how to control and exploit electronic and spin interactions and quantum fluctuations in bulk materials with novel functionality. Initially it was believed that quantum fluctuations can only be appreciable in magnetic materials where magnetic frustration is present,[7-11] however, it is now realized that strong spin-orbit coupling (SOC) can also produce the energies needed for quantum fluctuations.[12-14]

As the heaviest non-radioactive element, bismuth, in main group V, is located at the Zintl border in the periodic table.[15-20] Bi has moderate electronegativity (and can commonly act as either an anion or a cation) and strong spin-orbit coupling. Recent studies have shown that low dimensional materials with Bi layers are ideal platforms for realizing quantum topological electronic states, newly appreciated states of quantum matter.[21-23] The trigonal $La_2O_3$-type $Mg_3Bi_2$ has been proved to be an ideal platform for realizing type-II nodal line topological electronic states.[24] Usually, large magnetic moments on $Mn^{2+}$ can cause extremely large $d$-band splitting between spin-up and spin-down ($> 7$ eV), which drives Mn $d$ band far away from the Fermi level.[25] Some well-known intriguing behaviors in Mn-containing materials are giant magnetoresistance (GMR),[26-30] spin canting[31-33] and metal-insulator transition (MIT),[35-39] and thus here we focus on a previously studied material, $CaMn_2Bi_2$, which displays both magnetism and strong spin-orbit coupling.[40,41] $CaMn_2Bi_2$ crystalizes in the same structure as $Mg_3Bi_2$, a trigonal $La_2O_3$-type structure with the $P$-$3m$1 space group at ambient pressure (*not* in the "usual" tetragonal 122 structure), in which layers of Mn and Bi are separated by Ca layers.[42] The Mn is in a bilayer that can be considered as a puckered Mn honeycomb, which means that it has the potential to display unusual magnetically ordered states. The shortest Bi-Bi distance is ~4.65 Å and Mn resides in the center of $Bi_4$ tetrahedron. Bi atoms may host possible topological electronic states and $Mn@Bi_4$ tetrahedron offers the platform for the interplay between magnetic moment from Mn and SOC effect from Bi.[43-47] The magnetic structure of ambient pressure $CaMn_2Bi_2$ obtained by neutron diffraction has shown that nearest neighbor Mn atoms in the bilayer have opposing spins.[40] A common magnetic phenomenon associated with SOC is spin-canting, which has been observed and extensively investigated in various quantum systems such as $SrMnBi_2$ and $YbMnBi_2$.[48-50] Unfortunately, there is no direct evidence showing that spin-canting exists in $CaMn_2Bi_2$, suggesting that the interaction between SOC and magnetism is weak in this system at ambient pressure.

It is widely accepted that high pressure is the best tool for changing and controlling interatomic distances in solids – a fundamental property-determining parameter - because it cleanly changes such distances while not inducing structural disorder, which is the case for chemical pressure.[51-53] This tool has long been used, for example, to tune semiconductors, superconductors, and other solid-state materials to better understand their band structures and probe the electronic levels that arise from defects or impurities.[54-67] Pressure is particularly useful because the effects it induces are typically not associated with large changes in entropy, and therefore it frequently does not obscure subtle phase transitions. As a consequence, atomic and magnetic interaction parameters such as orbital energy and orbital overlap can be tuned and detected under pressure, and in the quantum material $CaMn_2Bi_2$ may potentially enhance the interplay between SOC and magnetism.

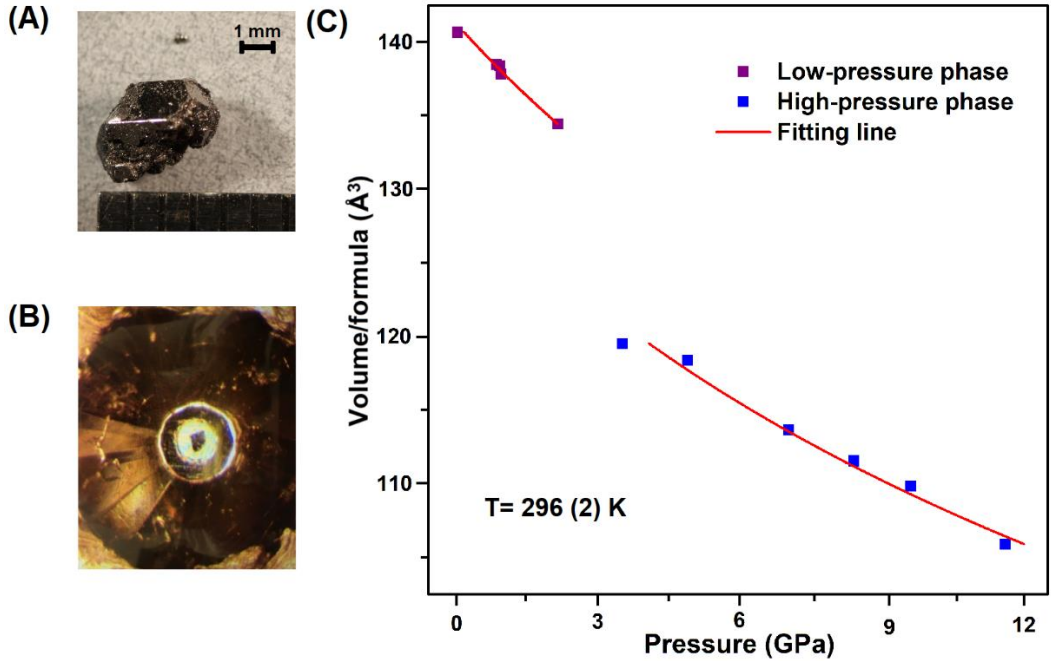

**FIG. 1.** **(A)** A $CaMn_2Bi_2$ single crystal. **(B)** Enlarged view of a Diamond Anvil Cell (DAC) filled with small $CaMn_2Bi_2$ single crystal and pressure medium. **(C)** Pressure-dependence of the volume per formula unit of $CaMn_2Bi_2$. The black squares, blue circles and red lines represent the low-pressure phase, the high-pressure phase and the equation of state fit lines, respectively.

Firstly, we performed the study on crystal structures of $CaMn_2Bi_2$ under various pressures. According to the results of single crystal X-ray diffraction experiments, up to 2.21 GPa $CaMn_2Bi_2$ maintains its previously reported ambient pressure trigonal crystal structure (in space group $P\text{-}3m1$).[40] The diffraction patterns show that by compression to 3.56 GPa the sample has undergone a first order displacive phase transition and the space group changes to monoclinic $P2_1/m$. Due to the transformation mechanism, the sample retains its single crystal character. The detailed crystallographic data, atomic coordinates and equivalent isotropic displacement parameters for $CaMn_2Bi_2$ at ambient pressure, 2.21 GPa and 11.6 GPa are listed in Tables S1 and S2. In order to further characterize the material under pressure, we plot the curves of volume per formula unit *vs.* pressure in FIG. 1C and fit the data using the second-order Birch-Murnaghan equation of state: $P(V) = \frac{3B_0}{2}\left[\left(\frac{V_0}{V}\right)^{\frac{7}{3}} - \left(\frac{V_0}{V}\right)^{\frac{5}{3}}\right]$, where P is the pressure, $V_0$ is the reference volume, V is the deformed volume and $B_0$ is the bulk modulus.[70] The phase transition is clearly visible in FIG. 1C as a ~ 7.5% discontinuity in the pressure dependence of the volume. The bulk moduli of low-pressure and high-pressure phases are fitted to be 38.6 (1.2) Pa and 36.1 (4.1) Pa, respectively. The unit cell's volume of the high-pressure phase at the transition pressure $V_0$ is obtained as 131.4 (1.9) Å$^3$. FIG. 2 shows both the low-pressure and high-pressure crystal structures of CaMn$_2$Bi$_2$. As can be seen in FIG. 2A (left), the ambient-pressure phase consists of slightly distorted edge-shared Ca@Bi$_6$ octahedra with a tilting angle of only 0.24(2)° between Ca-Mn-Ca chains and the axis of regular octahedra, which increases to 0.36 (1)° at 0.99 GPa and 0.58(1)° at 2.21 GPa. In the low-pressure phase, the Mn atoms form a puckered honeycomb when viewed down the *c*-axis (bottom right in FIG. 2A). The magnetic structure of this material, determined by ambient pressure neutron diffraction experiments, shows opposing magnetic moments alternating between the nearest neighboring Mn atoms.[40,41] The high-pressure phase displays similar structural characteristics for the Ca@Bi$_6$ octahedra, except for the larger tilt angle and the Ca-Bi atomic distances. All Ca-Bi bond lengths at ambient pressure are 3.266 (1) Å while they range from 2.97(10) to 3.15(14) Å under 11.6 GPa. The Mn net, in contrast, exhibits distinct changes under high pressure (FIG. 2B). Clearly, the Mn net rearranges from a strongly puckered honeycomb to parallel one-dimensional (1D) zig-zag ladder-like chains. The Mn-Mn separations in the ambient pressure trigonal phase are uniform at 3.238 Å while the near neighbor Mn-Mn distances within the change are much shorter, ranging from 2.69 (14) to 2.84 (10) Å in the monoclinic phase at 11.6 GPa. Importantly, the Mn-Mn distance between 1D chains is much longer, at 3.53(17) Å at 11.6 GPa.

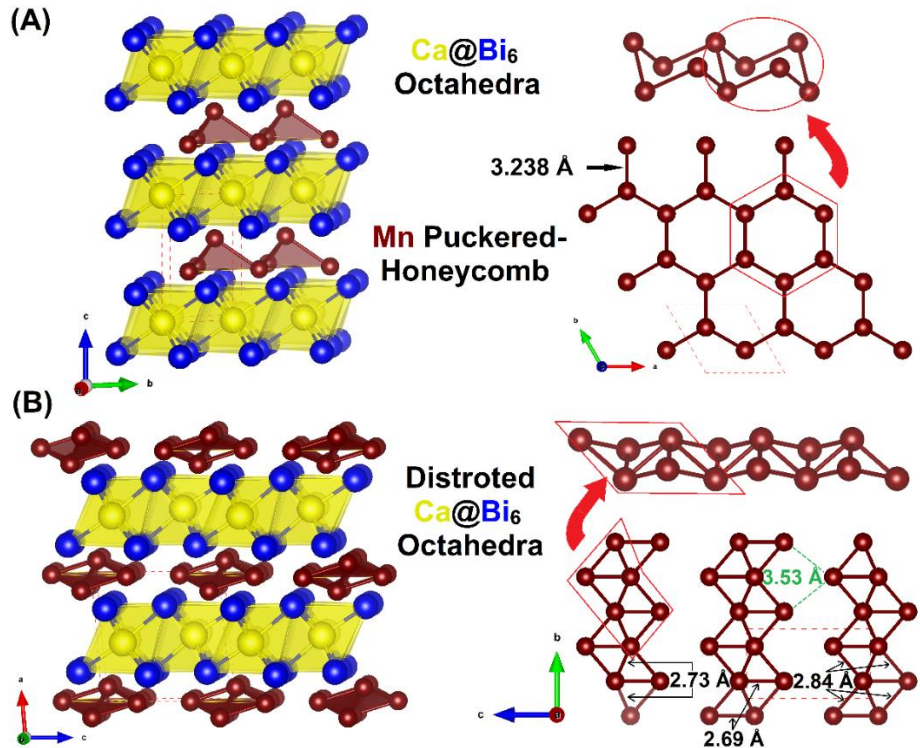

**FIG. 2.** **(A)** Low-pressure structure of CaMn$_2$Bi$_2$ where yellow, brown and blue balls represent Ca, Mn and Bi atoms. **(Left)** The overview of the crystal structure consisting with edge-shared Ca@Bi$_6$ octahedra and Mn-puckered honeycomb. **(Right)** The view on edge-shared Mn$_6$ ring in 3-dimensional (3D) network. **(B)** Crystal structure of CaMn$_2$Bi$_2$ under high pressure. **(Left)** The crystal structure consisting with distorted

edge-shared Ca@Bi$_6$ octahedra layers and distorted edge-shared Mn$_6$ chair conformation. **(Right)** The view on distorted edge-shared Mn$_6$ ring in quasi-one-dimensional (quasi-1D) chain.

We have also measured the high-pressure resistivity of CaMn$_2$Bi$_2$ single crystals to verify the pressure-induced structural transformation and to characterize the electronic properties of the phases under pressure. Two CaMn$_2$Bi$_2$ single crystals labeled as sample #1 and #2 were measured and excellent reproducibility of the results was found. FIG. 3A displays the pressure dependence of the resistivity ($\rho(P)$) for these two samples measured between 0-3 GPa at room temperature. The resistivity of CaMn$_2$Bi$_2$ drops suddenly by two orders of magnitude at $P_c$ = 2.15(5) and 2.35(5) GPa for samples #1 and #2, respectively, in good agreement with the structural phase transition detected by high-pressure single crystal XRD. The transition width of $\rho(P)$ around $P_c$ is less than 0.1 GPa, reflecting good pressure homogeneity in our measurements with the CAC and the dramatic nature of the phase transition. We note that the slight variation of $P_c$ values for the two samples likely arises from the different compression procedures employed. The critical pressure $P_c$ = 2.35(5) GPa observed for sample #2 is more accurate for which the reason is explained in the supporting information. We have measured the temperature dependence of resistivity $\rho(T)$ from room temperature down to 2 K at different pressures up to 7 GPa for these two samples. (FIG 3B.) At elevated temperatures (T>T$_N$), the curves exhibit "metallic" behavior originated from a strongly temperature dependent mobility, which is not metallic, investigated by ref. 40. The behavior below the structural phase transition at $P_c$ has the same characteristics as are observed at ambient pressure, with a kink anomaly at the antiferromagnetic transition $T_N$ ~150 K, followed by an upturn at low temperatures. The transition temperature $T_N$ can be defined clearly from the sharp peak in d$\rho$/d$T$ (FIG. 3B). With increasing pressure, both $T_N$ and the resistivity minimum gradually move to higher temperatures. Meanwhile, a resistivity plateau appears at low temperature and becomes more pronounced with increasing pressure. Given that the Mn-Mn interatomic distance decreases slightly with increasing pressure in the low pressure phase, these results suggest that the antiferromagnetic interactions between the Mn localized moments, mediated by the conduction electrons, are strengthened upon the pressure-induced decrease of the interatomic distances. The results for sample #1 are given in FIG. 3B on a semi-logarithmic scale. Interestingly, $\rho(T)$ at $P$ > $P_c$ displays some mild anomalies that vary systematically with pressure. For example, $\rho(T)$ at 2.5 GPa exhibits a kink-like anomaly at around 200 K followed by a shoulder-like anomaly centered around 116 K. These features can be described by a symmetric peak at $T_1$ and a broad hump at $T_2$ in the d$\rho$/d$T$ curve, as shown in the middle of FIG. 3B. With increasing pressure, $T_1$ shifts quickly to higher temperatures while $T_2$ moves to lower temperatures, and the magnitude of these anomalies in d$\rho$/d$T$ curve are progressively suppressed. FIG. S1 displays the $\rho(T)$ and d$\rho$/d$T$ curves at 3 and 5.5 GPa for sample #2, which shows nearly identical behavior as shown in FIG. 3B for sample #1. These anomalies can be seen most clearly in FIG. S1, where a linear scale of resistivity is employed. Based on the investigation of ref. 40, low-pressure phase of CaMn$_2$Bi$_2$ shows band gap at Fermi level within all tested temperature range while the resistivity measurement conducted under high pressure drops dramatically with 3 to 5 orders of magnitude which is an indication of insulator to metal transition. Moreover, the band structure calculated by LMTO shown in FIG.

S4B clearly presents a gapless feature for high-pressure phase which theoretically proves that the insulator to metal transition appears when $CaMn_2Bi_2$ is pressurized to $P_c$.

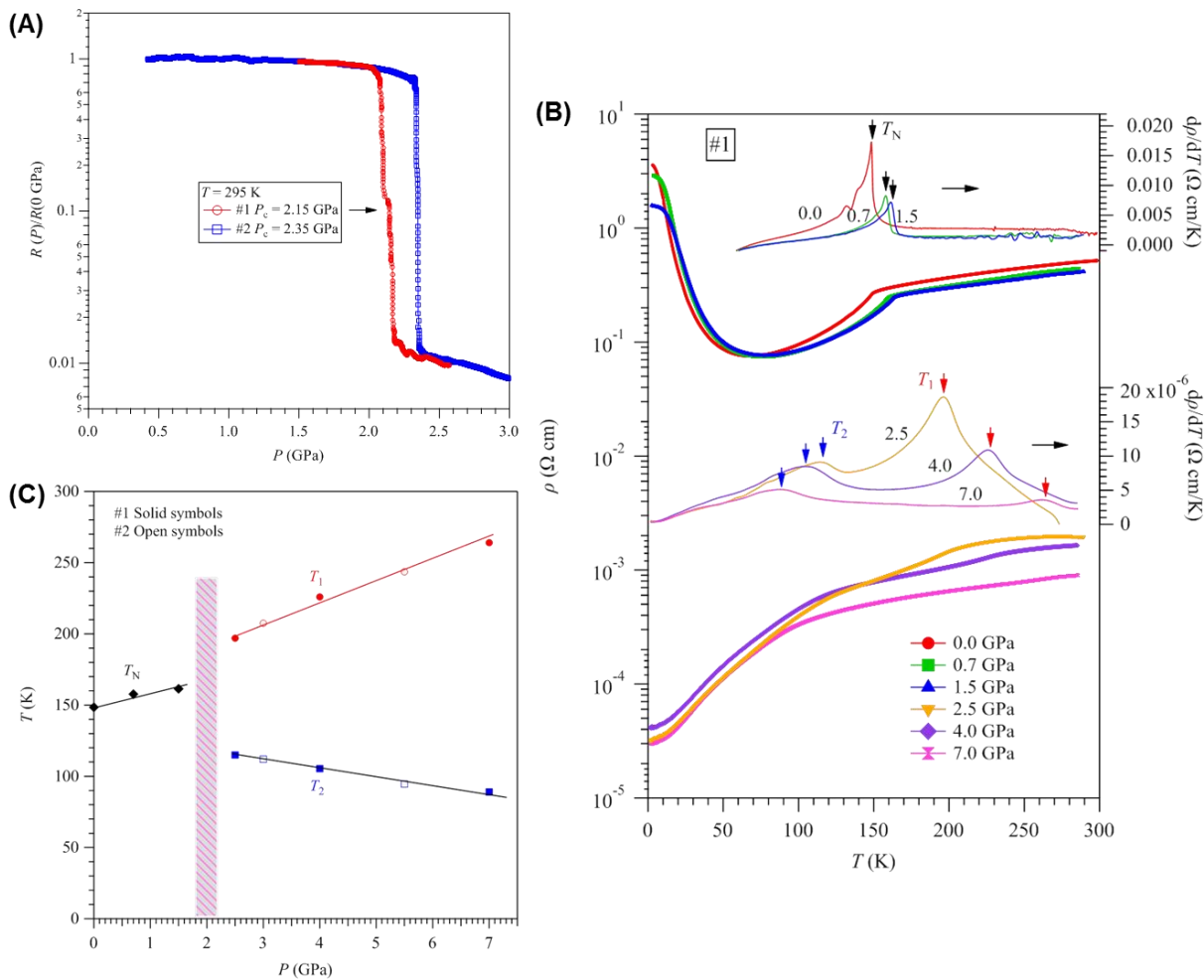

**FIG. 3 (A)** Pressure dependence of resistivity for two $CaMn_2Bi_2$ single crystals (samples #1 and #2) at room temperature. **(B)** Temperature dependence of resistivity $\rho(T)$ and its derivative $d\rho/dT$ under various pressures up to 7 GPa for a $CaMn_2Bi_2$ single crystal (sample #1). (C) Temperature-pressure phase diagram of the transitions in $CaMn_2Bi_2$ single crystals.

Based on the high-pressure resistivity measurements described above, we can construct FIG. 3C, the temperature-pressure phase diagram for $CaMn_2Bi_2$. The material undergoes its unique structural phase transition around $P_c \approx 2.35$ GPa, as described above. In the low-pressure trigonal phase, the antiferromagnetic transition temperature increases slightly under pressure. The high-pressure monoclinic phase, in contrast, is a metal that undergoes two phase transitions, at $T_1$ and $T_2$, which exhibit opposite pressure dependences within the pressure range investigated. Since the resistivity decreases with a larger slope at $T_1$ (FIG. 3B & S1), it is likely that the transition at $T_1$ represents long-range magnetic order (the scattering rate for conduction electrons is reduced when magnetic moments are ordered). This is supported by the fact that the positive pressure effect for $T_1$ is similar to that seen for $T_N$ in the low-pressure phase. In contrast, the broad hump feature around $T_2$ may be associated with the formation of a spin or charge density wave phase, which can be suppressed by pressure. This scenario is consistent with the chain-based quasi-1D structure and enhanced electrical conductivity in the high-pressure phase. Further neutron diffraction study would be of interest to determine the origin of these transitions in high-pressure chain-based $CaMn_2Bi_2$.

In summary, we have investigated the compression dependence of the crystal structures and electric properties of $CaMn_2Bi_2$. Our high-pressure single crystal X-ray diffraction studies show that $CaMn_2Bi_2$ undergoes a structural phase transition between 2 and 3 GPa and assumes a unique chain-based monoclinic structure with space group *P*2$_1$/*m* at high pressure. The isolated 1D Mn chains can host complex magnetic ordering, which has been modeled in the supporting materials. Electrical resistivity measurements at pressures below the structural phase transition indicate that pressure increases the antiferromagnetic transition temperature. The high-pressure monoclinic phase exhibits metallic properties with two distinct temperature-driven phase transitions. The positive pressure effect of one of these phase transitions is similar to the pressure effect on $T_N$ in the low-pressure phase, reflecting that it is likely due to antiferromagnetic ordering. The negative pressure effect displayed by the second electronic transition suggests that it may be related to the formation of a spin or charge density wave phase, resulting from a Fermi surface instability of the quasi-1D Mn chains. These results demonstrate that pressure provides a versatile tool for the characterization of magnetic topological quantum materials such as $CaMn_2Bi_2$.

## *Supporting Information*

The Supporting Information is available free of charge on the ACS Publications website at DOI: xxxxx.

Sample Preparation of $CaMn_2Bi_2$; Single Crystal Structure Determination at Ambient Pressure; Single Crystal Structure Determination at High Pressure; Single Crystal Structure Analysis; Electrical Measurements at Various Pressures; Electronic Structural Calculations and Analysis.

## *Author Information*

Corresponding Author: Dr. Weiwei Xie (weiweix@lsu.edu) and Dr. Jinguang Cheng (jgcheng@iphy.ac.cn)

Notes: The authors declare no competing financial interest.

## *Acknowledgement*

The work at LSU was supported by the U.S. Department of Energy, Office of Science, Basic Energy Sciences, under EPSCoR Grant No. DE-SC0012432 with additional support from the Louisiana Board of Regents. Authors deeply appreciate technical support from Louisiana State University-Shared Institute Facility (SIF) for SEM-EDS. G.F. and P.D. acknowledge support from the University of Hawaii Materials Science Consortium for Research and Education (MSCoRE).

JGC is supported by the National Key R&D Program of China (2018YFA0305702), the National Natural Science Foundation of China (11574377, 11834016, and 11874400), the Strategic Priority Research Program and Key Research Program of Frontier Sciences of the Chinese Academy of Sciences (XDB25000000 and QYZDB-SSW-SLH013).

# Pressure-Induced Large Volume Collapse, Plane-to-Chain, Insulator to Metal Transition in CaMn$_2$Bi$_2$

Xin Gui,[a] Gregory J. Finkelstein,[b] Keyu Chen,[cd] Tommy Yong,[b] Przemyslaw Dera,[b] Jinguang Cheng,[cde*] Weiwei Xie [a*]

[a] Department of Chemistry, Louisiana State University, Baton Rouge, LA, USA 70803
[b] Hawai'i Institute of Geophysics and Planetology, University of Hawai'i at Manoa, Honolulu, HI, USA 96822
[c] Beijing National Laboratory for Condensed Matter Physics and Institute of Physics, Chinese Academy of Sciences, Beijing, China 100190
[d] School of Physical Sciences, University of Chinese Academy of Sciences, Beijing, China 100190
[e] Songshan Lake Materials Laboratory, Dongguan, Guangdong, China 523808

**General Experimental Methods**

**Sample Preparation of $CaMn_2Bi_2$:** For single crystals of $CaMn_2Bi_2$, the Bi self-flux method was employed by mixing Ca, Mn, and Bi with a molar ratio of 1:2:10 in an alumina crucible sealed in a quartz tube. The tube was heated to 1000 °C at a rate of 180 °C/h and held there for 48 hours. After that, the sample was slowly cooled down to 400 °C at a rate of 6 °C/h. Hexagonal-shape single crystals (~3×3×3 $mm^3$, e.g. FIG. 1A) of $CaMn_2Bi_2$ were obtained after removing excess Bi by centrifuging.

**Single Crystal Structure Determination at Ambient Pressure:** Single crystal data under ambient conditions was collected on a Bruker Smart Apex II diffractometer equipped with Mo $K_\alpha$ ($\lambda$=0.71073 Å) radiation. Samples mounted on a Kapton loop were measured with a scan angle of 2° and an exposure time of 10 seconds. The structure was solved using direct methods and refined using full-matrix least-squares on $F^2$, as implemented in SHELXTL packages.[1] Peak intensity corrections were carried out by Bruker SMART software with numerical absorption correction using face-indexed model in XPREP.[2]

**Single Crystal Structure Determination at High Pressure:** High pressure experiments were carried out at the X-ray Atlas Lab at the Hawaii Institute of Geophysics and Planetology. A piece of the same single crystal that was analyzed at ambient pressure was mounted inside a diamond anvil cell (DAC) with 300-μm culet diamonds. The microphotograph of the sample inside the DAC is shown in FIG. 1B. The rhenium gasket was pre-indented to ~50 μm prior to sample loading. A ~200-μm hole was cut in the gasket indentation using a 100W 1064 nm YAG laser. Cubic boron nitride *c*BN (upstream) and tungsten carbide WC (downstream) backing plates were used. A mixture of methanol, ethanol and water in a 16:3:1 volume ratio was used as a pressure transmitting medium. The sample was loaded into the gasket hole along with two small ruby spheres placed on the opposite sites of the crystal, used for in situ pressure determination.[3] The high-pressure diffraction measurements were carried out on a customized Bruker D8 Venture diffractometer with a Phonon II detector, fixed-chi goniometer and Incoatec IμS 3.0 Ag K$\alpha$ microfocus source with Helios focusing optics. The diamond anvil cell was mounted on a motorized XYZ stage attached directly to the omega platform and centered using radiographic scans with a photodiode detector, similar to the procedure used at synchrotron high-pressure beamlines. Data collection was carried out using only ω-scans with 11° steps and an exposure time of 20 seconds per frame. Eight different scans were collected to reach a higher completeness. Pressure was increased gradually from 0 to 11.6 GPa and diffraction data was analyzed with the Bruker APEX 3 software. The structure of the high-pressure phase was solved using the intrinsic phasing algorithm.[1] The single crystal refinement was done using the SHELXTL package with full-matrix least-squares on $F^2$.[1]

**Table S1.** Crystallographic data obtained from single crystal X-ray diffraction of CaMn$_2$Bi$_2$ at 296(2) K under various pressures.

| Pressure (GPa) | AP | 2.21 | 11.6 |
|---|---|---|---|
| Space group; Z | P-3m1; 1 | P-3m1; 1 | P2$_1$/m; 2 |
| a(Å) | 4.629 (1) | 4.5498 (2) | 7.212 (4) |
| b(Å) | 4.629 (1) | 4.5498 (2) | 4.060 (1) |
| c(Å) | 7.628 (2) | 7.5000 (5) | 7.260 (6) |
| β (°) | 90 | 90 | 94.85 (7) |
| V (Å$^3$) | 141.54 (9) | 134.46 (2) | 211.8 (2) |
| Extinction Coefficient | 0.025 (3) | 0.014 (5) | 0.000 (7) |
| θ range (deg) | 2.670 – 33.306 | 4.081 – 27.088 | 3.017 – 23.450 |
| No. reflections; $R_{int}$ | 2062; 0.0725 | 797; 0.0241 | 780; 0.0580 |
| No. independent reflections | 245 | 115 | 120 |
| No. parameters | 10 | 10 | 17 |
| $R_1$: $\omega R_2$ (all $I$) | 0.0242; 0.0506 | 0.0173; 0.0316 | 0.1366; 0.3192 |
| Goodness of fit | 1.149 | 1.614 | 1.298 |
| Diffraction peak and hole (e$^-$/Å$^3$) | 2.929; -1.524 | 1.518; -2.054 | 3.670; -4.434 |

**Table S2.** Atomic coordinates and equivalent isotropic displacement parameters of CaMn$_2$Bi$_2$ system under different pressures ($U_{eq}$ is defined as one-third of the trace of the orthogonalized $U_{ij}$ tensor (Å$^2$)).

| Pressure | Atom | Wyckoff. | Occ. | x | y | z | $U_{eq}$ |
|---|---|---|---|---|---|---|---|
| AP | Bi | 2d | 1 | 1/3 | 2/3 | 0.2462 (1) | 0.0109 (2) |
| | Mn | 2d | 1 | 1/3 | 2/3 | 0.6198 (2) | 0.0136 (3) |
| | Ca | 1a | 1 | 0 | 0 | 0 | 0.0142 (6) |
| 2.21 GPa | Bi | 2d | 1 | 1/3 | 2/3 | 0.2439 (1) | 0.0125 (2) |
| | Mn | 2d | 1 | 1/3 | 2/3 | 0.6190 (3) | 0.0149 (4) |
| | Ca | 1a | 1 | 0 | 0 | 0 | 0.0154 (7) |
| 11.6 GPa | Bi | 2e | 1 | 0.754 (3) | ¼ | 0.133 (4) | 0.017 (3) |
| | Bi | 2e | 1 | 0.233 (3) | ¼ | 0.420 (4) | 0.023 (3) |
| | Mn | 2e | 1 | 0.115 (10) | ¼ | 0.062 (14) | 0.017 (7) |
| | Mn | 2e | 1 | 0.001 (10) | ¼ | 0.699 (14) | 0.015 (7) |
| | Ca | 2e | 1 | 0.523 (16) | ¼ | 0.770 (2) | 0.027 (12) |

**Electrical Transport Measurements under High Pressure:** The high-pressure resistivity of a single crystal of $CaMn_2Bi_2$ was measured with a standard four-probe method by using the palm cubic anvil cell (CAC) apparatus at the Institute of Physics, CAS.[4] Details about the experimental setup can be found elsewhere.[4] Glycerol was employed as the pressure transmitting medium and the pressure values were calibrated at room temperature by observing the characteristic phase transitions of elemental bismuth.[5,6] For our measurements with the CAC, we always changed the pressures at room temperature, tightened the pressure cell at the target pressure, and then loaded it into a liquid-helium cryostat to measure the temperature dependence of the resistivity at a given pressure. For sample #1, we tightened the CAC at 0.7 and 1.5 GPa and then measured the $\rho(T)$ curves from room temperature down to 2 K at these two pressures, as shown in the upper part of FIG. 3B. Afterwards, we observed the phase transition when increasing pressure from 1.5 to 2.5 GPa at room temperature. For this procedure the actual pressure experienced by sample #1 should be slightly higher than that shown in FIG. 3A. To more accurately detect the pressure for the phase transition, we continuously applied pressure from 0 to 3 GPa at room temperature for sample #2, using a procedure similar to what we used for the bismuth pressure calibration.

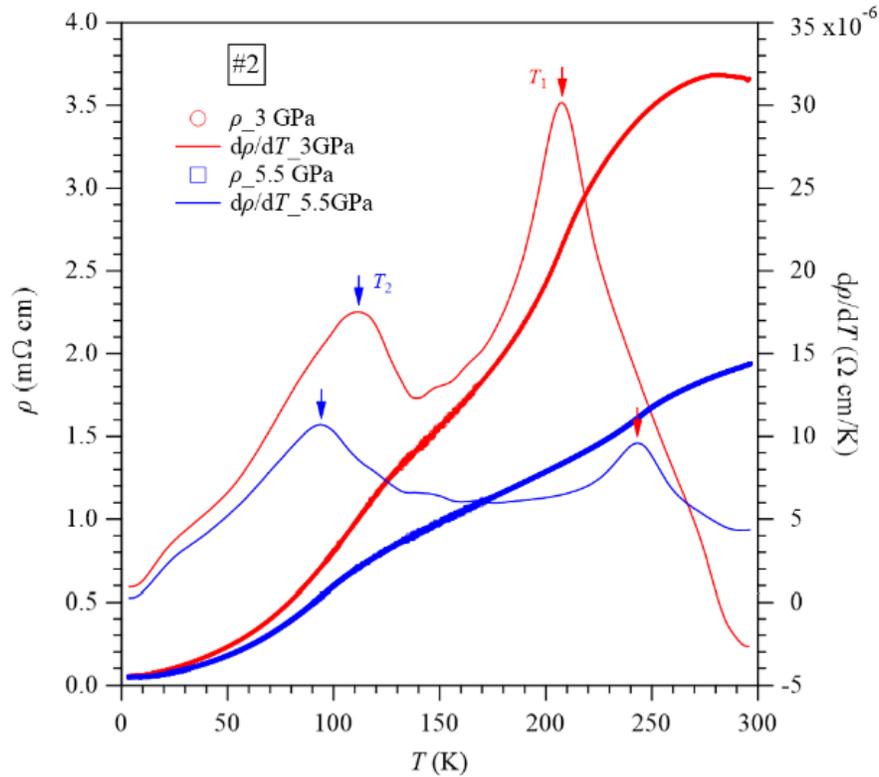

**Figure S1.** Temperature dependence of resistivity $\rho(T)$ and its derivative $d\rho/dT$ at 3 and 5.5 GPa for sample #2.

**Electronic Structure Calculations for $CaMn_2Bi_2$:** Based on the experimental crystal and magnetic structure of $CaMn_2Bi_2$ at ambient pressure, the electronic structure was calculated using WIEN2k with spin-orbit coupling (SOC) included.[7] Spin polarization using the Local Density Approximation (LDA) was employed on the Mn atoms for the characterization of the antiferromagnetically ordered phase.[8] The energy cutoff was 500 meV. Reciprocal space integrations were completed over an 8×8×4 Monkhorst-Pack *k*-point mesh with the linear tetrahedron method.[9] With these settings, the calculated total energy converged to less than 0.1 meV per atom. We also employed Tight-Binding, Linear Muffin-Tin-Orbital Atomic-Spheres-Approximation (TB-LMTO-ASA) calculations to characterize the band structure and density of states (DOS) for the high-pressure phase.[10] A mesh of 64 *k* points was utilized while 0.05 meV was set as the convergence criterion. We filled the empty space inside the unit cell with overlapping Wigner-Seitz (WS) spheres with a combined correction for the overlapping parts.[11] The WS radii are: 1.780 Å for Ca; 1.530 Å for Mn; and 1.670 Å for Bi. The overlap of WS spheres was limited to no larger than 16%. The basis set for the calculations included Ca 4*s*, 3*d*; Mn 4*s*, 4*p*, 3*d*; and Bi 6*s*, 6*p* wavefunctions.

**Electronic Structure and Total Energy Calculations for the $CaMn_2Bi_2$ phases under pressure.** To simulate the transition pressure, a series of total energy calculations were performed for both the low- and high-pressure phases by the LMTO method. Starting with the lattice parameters we determined under ambient pressure (for low-pressure phase) and 11.6 GPa (for high-pressure phase), also with decrements of -2% and -1% and increments of 0.5%, 1%, 1.5%, 2%, 2.5%, 3%, 4%, 5% for low-pressure phase and increments of 1%, 1.5%, 2%, ……, 7.5%, 8.5%, 9%, 9.5%, 10% for high-pressure phase, we compress/expand the unit cell with fixed *c*/*a* ratio for low-pressure phase and *c*/*a* & *b*/*a* for high-pressure phase and the value of β for high-pressure monoclinic phase. In FIG. S2, we plot the energy/atom *vs* the volume/atom, with both curves fitted by the Birch-Murnaghan equation of state as described above. The calculations show that the high-pressure $CaMn_2Bi_2$ structure has lower energy over the whole calculated volume range, never displaying higher energy than the low-pressure phase. This indicates that according to LMTO calculations, the new high-pressure phase is more energetically stable than the ambient pressure phase. The reason why simulated EV curves do not show a energy-equivalent point, which indicates the exact volume/atom that phase transition takes place, may due to the isotropic expansion or compression we used in the simulation while it behaves anisotropic in the experiments. The band structure calculated for the ambient pressure phase using WIEN2k, shown in FIG. S3, is similar to that reported previously, showing semiconducting character with an indirect band gap ~ 0.05 eV.

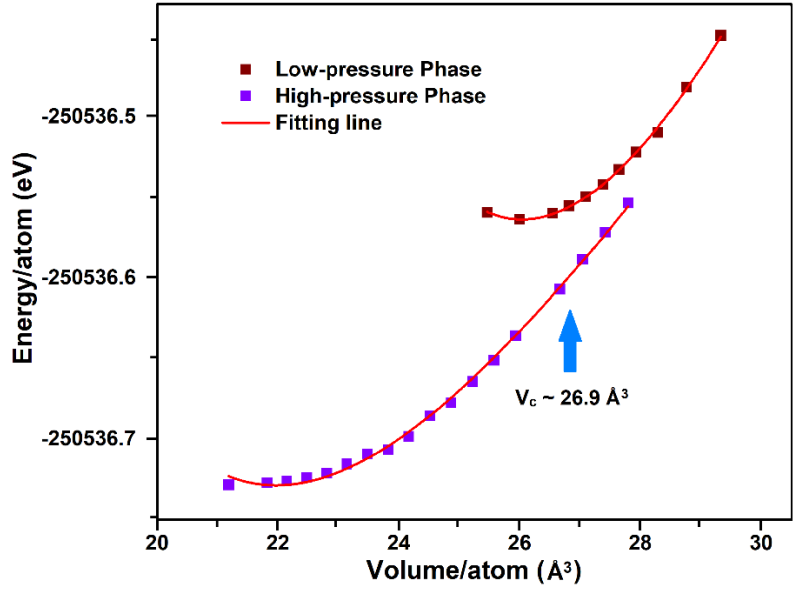

**FIG. S2.** Total energy calculation for low- & high-pressure phases of $CaMn_2Bi_2$. The brown and purple squares represent the total energy per atom for low- and high-pressure phases. The red fitting lines are fitted by Birch–Murnaghan equation of state. The approximate volume near critical pressure has been marked as $V_c$.

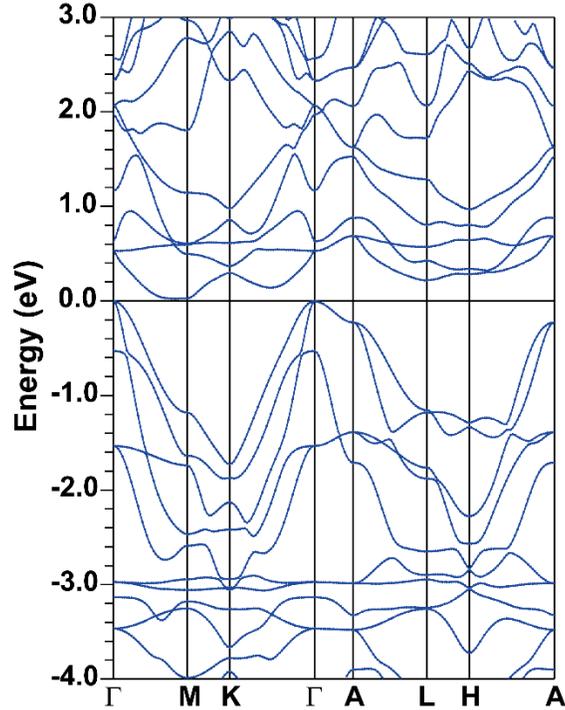

**FIG. S3.** Band structure calculated for $CaMn_2Bi_2$ under ambient pressure with spin-orbit coupling and antiferromagnetic ordering.

We calculated the band structure and DOS to get more insight into the structural stability and magnetic properties of chain-based monoclinic $CaMn_2Bi_2$ under high pressure. To examine the possible magnetic structures, first-principles calculations were employed to evaluate the total energies and magnetic moments for five different cases, including non-magnetic (NM), ferromagnetic (FM), and three different hypothetical antiferromagnetic phases (AFM1, AFM2, and AFM3). All three possibilities for the antiferromagnetic phases are illustrated in FIG. S4 A, where the red and orange arrows indicate opposite magnetic moments, but not fixed spin orientations. Among the five different cases, AFM2 has the lowest calculated total energy, indicating that AFM2 is the most stable magnetic arrangement among the four hypothetical magnetic models for the high-pressure system. In the AFM2 magnetic ordering model, the Mn moment is 0.0112 emu/formula which indicates a possibility of spin-canting.

Based on our determination of the high-pressure crystal structure, monoclinic $CaMn_2Bi_2$ consists of individual Mn chains, which can be recognized as a network of edge-sharing distorted $Mn_6$ clusters in chair conformations. The $Mn_6$ chair cluster is analogous to what is observed in trigonal $CaMn_2Bi_2$ at ambient pressure. In trigonal $CaMn_2Bi_2$, the magnetic moments are oriented antiparallel in the two layers of the puckered honeycomb. The antiparallel oriented magnetic ordering in AFM2 shown in FIG. S4 A is similar to the magnetic ordering observed in trigonal $CaMn_2Bi_2$. The total uncompensated magnetic moment of the AFM2 model is determined to be 0.0112 $\mu_B$ per unit (i.e. greater than zero), which may result from the occurrence of spin-canting induced ferrimagnetism for $CaMn_2Bi_2$ at high pressure.

The band structure and DOS results for the high-pressure phase with anti-ferromagnetic ordering are shown in FIG. S4 B. (For reference, the electronic structures for NM (FM), AFM1 & AFM3 models are shown in FIG. S5.) Most of the DOS curves between -3.5 and +1.5 eV are derived from the Mn 3$d$ band; below -3 eV is a ~1 eV tail comprising a combination of Mn and Bi 6$s$ and 6$p$ orbital contributions. In the LDA-DOS curve, the Mn 3$d$ band exhibits little fine structure except for a sharp, intense peak, which is ~0.1 eV wide, just above 0 eV. According to the corresponding band structure curves, it is mostly the wave functions of the Mn 3$d$ orbitals that contribute to this peak. Moreover, the LDA band structure reveals that the peak involves Mn 3$d$ bands that are nearly dispersionless near the Brillouin zone boundaries.

Thus, according to the LDA-DOS curves, monoclinic chain-based $CaMn_2Bi_2$ at high pressure is susceptible either to a structural distortion by disrupting the strong Mn-Mn orbital interactions at the Fermi level or to magnetism by breaking the spin degeneracy, both of which will substantially decrease the density of electronic states at the Fermi Energy. Applying spin polarization via LSDA splits the DOS curves for the spin-up and spin-down wave functions, as seen in FIG. S4 B. The corresponding Fermi levels are shifted away from the peaks in the DOS curves and closely approach the noticeable pseudogap in the DOS curve. Analysis of the local magnetic moments reveals that the magnetic moments primarily arise from Mn atoms, corresponding to 2.21 µB per Mn atom (which nearly cancel in the AFM2 model). These calculated magnetic moments per Mn are significantly smaller than those at ambient pressure, which may be due to the strong spin orbit coupling effect of the Bi atoms. The electronic structures for all the hypothetical magnetic models lead to a typical metallic band structure for the high-pressure monoclinic phase, despite its antiferromagnetic character, consistent with the electrical resistivity measurements.

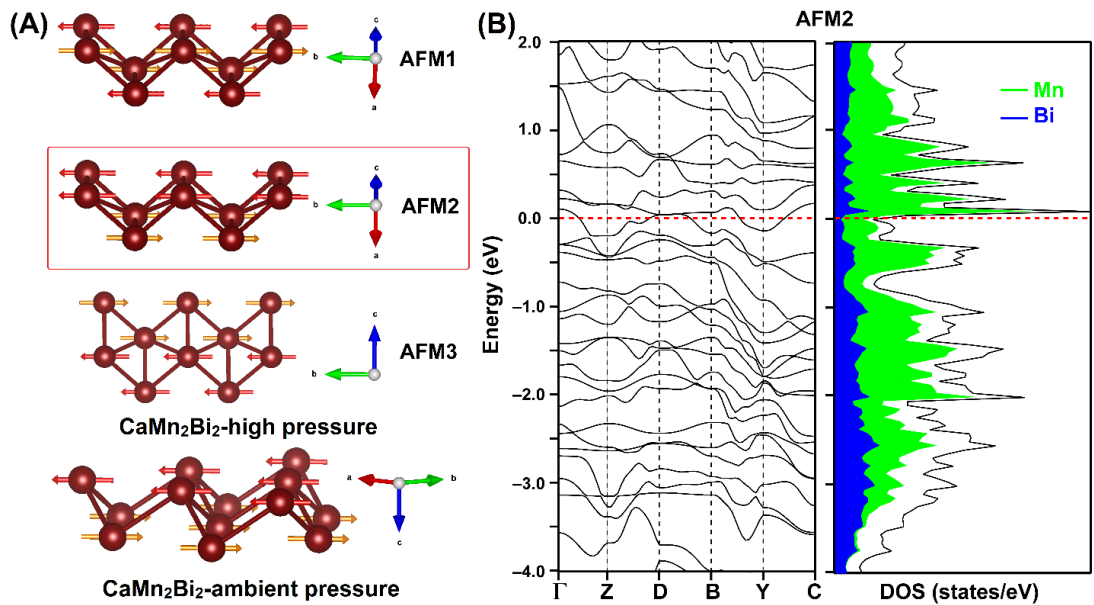

**Figure S4. (A)** Hypotheses of different magnetic orderings and magnetic structure under high pressure for $CaMn_2Bi_2$ where the red and orange vectors represent magnetic moments with opposite orientation. **(B)** The band structure and DOS of high-pressure $CaMn_2Bi_2$ with AFM2 magnetic ordering.

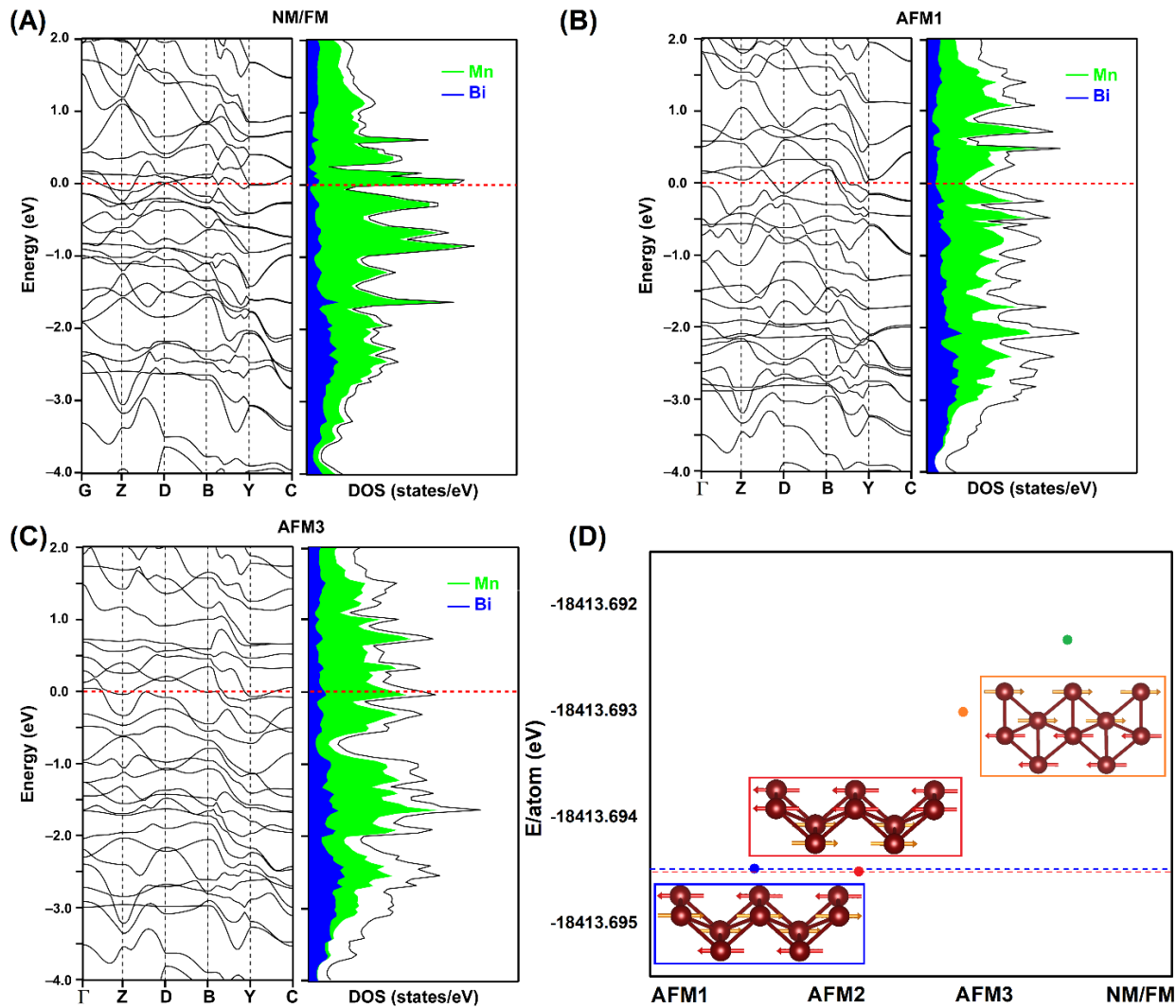

**Figure S5.** Band structure and density of states (DOS) of **(A)** non-magnetic (NM) (ferromagnetic (FM)), **(B) & (C)** antiferromagnetic (AFM1 & AFM3) for high-pressure CaMn$_2$Bi$_2$. **(D)** Total energy for different hypotheses.

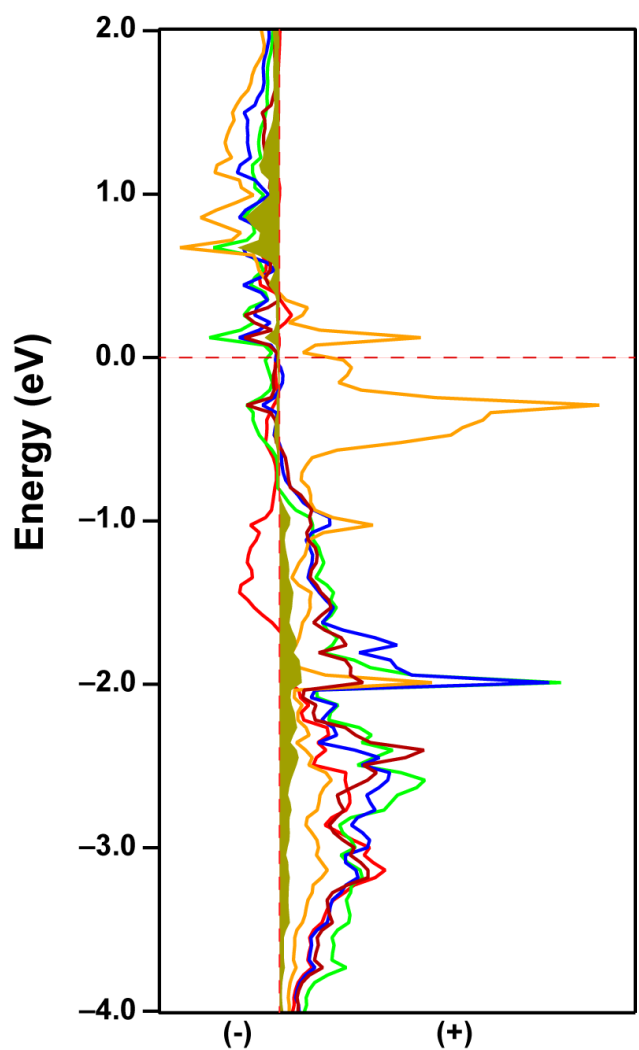

**FIG. S6.** Crystal Orbital Hamiltonian Population (-COHP) calculation of Mn-Mn interatomic interaction for high-pressure $CaMn_2Bi_2$. The curve between 1D chains (bond length=3.53 Å) is filled with olive green to show the weak interaction.